# Multivariate Analysis of Gut Microbiota Composition and Prevalence of Gastric Cancer

Aadhith Shankarnarayan
Dept of Computer Science Engineering
American University of Sharjah
Sharjah, United Arab Emirates
b00089801@aus.edu

Dheeman Gangopadhyay
Dept of Computer Science Engineering
American University of Sharjah
Sharjah, United Arab Emirates
b00092265@aus.edu

Dr. Ayman AlZaatreh
Dept of Statistics
American University of Sharjah
Sharjah, United Arab Emirates
aalzaatreh@aus.edu

*Abstract*— The global surge in the cases of gastric cancer has prompted an investigation into the potential of gut microbiota as a predictive marker for the disease. The alterations in gut diversity are suspected to be associated with an elevated risk of gastric cancer. This paper delves into finding the correlation between gut microbiota and gastric cancer, focusing on patients who have undergone total and subtotal gastrectomy. Utilizing data mining and statistical learning methods, an analysis was conducted on 16S-RNA sequenced genes obtained from 96 participants with the aim of identifying specific genera of gut microbiota associated with gastric cancer. The study reveals several prominent bacterial genera that could potentially serve as biomarkers assessing the risk of gastric cancer. These findings offer a pathway for early risk assessment and precautionary measures in the diagnosis of gastric cancer. The intricate mechanisms through which these gut microbiotas influence gastric cancer progression warrant further investigation. This research significantly aims to contribute to the growing understanding of the gut-cancer axis and its implications in disease prediction and prevention.

## I. INTRODUCTION

Gastric cancer is an incurable chronic disease affecting the colon and stomach. According to the European Cancer Journal, "Gastric cancer is the second most common cause of cancer mortality worldwide and remains a major global public health problem. After more than 30 years of clinical research, the prognosis for advanced gastric cancer (AGC) remains grim, with median overall survival (OS) barely exceeding 1 year until recently" [1]. Stomach cancer develops when cells in any part of the stomach grow and divide abnormally. Tumors can begin anywhere in a stomach, but most start in the glandular tissue on the stomach's inner surface [2]. Signs and symptoms of stomach cancer may include: trouble swallowing, belly pain, feeling bloated after eating, feeling full after eating small amounts of food, not feeling hungry when you would expect to be hungry, heartburn and indigestion.

Gastric cancer may not always show symptoms in its early stages. However, when symptoms do emerge, they may include indigestion and discomfort in the upper abdomen. Symptoms of gastric cancer may not become evident until the disease has progressed significantly. In the advanced stages of stomach cancer, individuals may experience symptoms such as profound fatigue, unexplained weight loss, hematemesis (vomiting blood), and melena (black, tarry stools) [3]. There are various treatment options for gastric cancer depending on many crucial factors, such as the stage of the cancer, your overall health, and your preferences. Occasionally, collaboration between gastroenterologists and cancer care teams occurs to devise a comprehensive treatment strategy, which may encompass a combination of therapeutic modalities. Some popular treatments include endoscopic mucosal resection, surgery (total gastrectomy and subtotal gastrectomy), radiation therapy, and chemotherapy [4] However, these mainstream treatments are prone to various side effects and can tremendously lower the patient's quality of life during the course of the treatment. It is critical to find alternative remedies to alleviate symptoms in a non-invasive manner that doesn't deteriorate the patient's quality of life. The first step would be exploring the potential gut-cancer link by finding taxa of gut bacteria associated with this disease. Interestingly, Zhang et al. assert that colorectal and gastric cancers, the major gastrointestinal tract cancers, are closely linked with the gut microbiome. Despite this assertion, the precise gut microbiota composition that correlates with gastric cancer remains elusive. Notably, in their studies, Zhang and colleagues uncovered significant shifts in gut microbiota composition and diversity among patients with gastric cancer, compared to a healthy subset of individuals [5]. Building upon prior research, this study aims to identify bacterial genera within families, focusing on the significant relationships between metabolites and gut bacteria observed in patients with gastric cancer and other gastrointestinal health conditions such as inflammatory bowel disease (IBD).

## II. DATASET AND PARTICIPANTS

This study is centered on analyzing microbiome data associated with gastric cancer and identifying genes potentially implicated in its development. The primary dataset used in this study, labeled "genera.csv," is derived from a comprehensive resource integrating meta-analyses across diverse disease contexts [6]. This particular dataset was selected for its emphasis on the microbiome biodata associated with gastric cancer.

The dataset "genera.csv" contains data from 106 participants who had undergone colonoscopy at the National Cancer Center Hospital of Tokyo, 56 participants who did not exhibit any colorectal findings or prior history of gastroenterological surgery were included as controls. The remaining 50 had undergone gastrectomy prior to the colonoscopy. Within this group, individuals without recent disease activity such as precancerous lesions, adenomatous polyps, or carcinomas were selected for inclusion in the dataset. However, 10 participants exhibited cancerous activity after the colonoscopy, leading to their exclusion from the dataset. After orally administrating the bowel cleansing agent at the hospital on the day of colonoscopy, the faecal samples were collected during the first defaecation [7].

The dataset is a genus-level abundance table comprising 16S-RNA sequenced metagenomic profiles from the fecal samples of 96 individuals, encompassing participants between 40 and 79 years of age. It features a detailed enumeration of 10,528 bacterial genera, each represented by a column indicating the relative percentage of that genus present in a participant's gut microbiome. The key attribute of this dataset is the 'Sample' column, which includes patient identifiers and their health status, categorized into two distinct classes: "Healthy" and "Gastrectomy." The latter refers to individuals who had undergone subtotal or total gastrectomy procedures as a treatment for gastric cancer.

The analysis aims to dissect the intricate relationships between various bacterial genera and the prevalence of gastric cancer, particularly in post-gastrectomy patients. Leveraging this rich dataset, the study seeks to uncover patterns and potential biomarkers that could significantly advance the understanding and management of gastric cancer.

## III. METHODOLOGY

Prior to implementing crucial feature selection methods, the data was preprocessed by removing low variance genera. Additionally, features exhibiting high correlation were analyzed, and within each pair of correlated features, the feature showing lower average correlation with other features was removed. Subsequently, the identified genera were effectively normalized to map the contributions of low-concentration genera. Following these preprocessing steps, the subsequent statistical methods were implemented:

### A. Information Gain Analysis

Information gain measures the effectiveness of an attribute in classifying training data. It can be used to select those features in a dataset that are most useful for predicting a target variable, thus reducing dimensionality and simplifying models. Sturges' rule is a method for determining the number of bins in a histogram [8]. This rule helps decide how to discretize a continuous attribute into different bins before applying information gain analysis.

### B. Least Absolute Shrinkage and Selection Operator (LASSO)

Lasso regression is a useful method for regularization and feature selection using a shrinkage method. It effectively compresses less significant features to zero by penalizing the sum of absolute coefficients, which enhances model interpretability and decreases overfitting [9]. It generates models that are easy to interpret, such as subset selection, while also demonstrating the stability seen in ridge regression.

### C. Linear Discriminate Analysis (LDA)

LDA is used to find a linear combination of features that separates two or more classes of objects or events. It is widely used for dimensionality reduction and pattern classification. The resulting combination may be used as a linear classifier, or more commonly, for dimensionality reduction before later classification [10].

### D. Leave-One-Out-Cross-Validation (LOOCV)

LOOCV is a validation technique used to assess the performance of machine learning models, especially in scenarios with limited data samples [11]. It is a variant of the K-Fold Cross-Validation where K equals the total number of subsets in the dataset, n. In LOOCV, the dataset is divided into n folds, with each fold containing a subset of the dataset. The model is then trained n times, with each iteration training on n-1 folds and leaving out one fold as the test set. This process iterates through each of the n folds ensuring that each fold serves as the test set only once. The average of the results from each iteration provides an estimation of the performance of the model. This method is valuable for model selection, parameter tuning, prevention of overfitting, and provision of an unbiased estimate of the model's generalization on unseen data.

### E. Principal Component Analysis (PCA)

PCA is a dimensionality reduction technique used to transform high-dimensional data into a lower-dimensional space while retaining as much variance as possible [12]. The process involves identifying principal components, which are linear combinations of the original features. The process involves standardizing data, computing the covariance matrix, eigen-decomposition to identify eigenvalues and eigenvectors, and selecting principal components based on their contribution to variance. This ordered selection creates an efficient lower-dimensional representation, facilitating meaningful analysis of complex datasets.

## IV. ANALYSIS

The dataset, characterized by its robustness and richness in diverse microbiotas, necessitates meticulous feature selection to analyze and condense the genera. These selected genera can potentially provide insights into determining whether a patient may have gastric cancer.

The comprehensive analysis of the project is structured in three main phases. Initially, emphasis is placed on the preparation and pre-processing of the dataset, which is critical for ensuring data quality and suitability for subsequent analysis. Following this stage, various feature selection techniques are employed to shrink the number of variables and retain only the most important genes. Finally, the core analysis of the data is conducted, utilizing classification techniques to validate the results obtained.

### A. Data Preparation and Preprocessing

The dataset mentioned earlier comprises 96 samples, each with the respective percentage concentration of 10,528 bacterial genera. Before commencing data analysis, effective preprocessing is vital. The pre-processing approach began with the removal of near-zero-variance variables, which exhibit minimal to no variation in their values and can thus be excluded from the analysis to enhance the performance of the model. To identify these variables, the standard deviation for all the samples and their respective genera concentrations was calculated. Through experimentation, a frequency cutoff ratio of 3 was determined, flagging genera with frequencies lower than this threshold as candidates for removal. Lastly to ensure accurate analysis and modeling, all genera concentrations were normalized within the range [0,1].

## B. Feature Selection of Bacterium Genes

### 1) Correlation Matrix and Binning for Information Gain

Initially, a Pearson correlation matrix was applied to identify and eliminate highly correlated variables, with a threshold set at 0.8 to determine strong correlations between pairs of features. Features exceeding this correlation threshold were systematically assessed. Within each highly correlated pair, the feature that exhibited a lower average correlation with other features in the dataset was flagged for removal. As a result, 4237 bacterial genera were retained. Subsequently, the concentrations of retained bacterial genera were binned according to Sturges' rule. Information Gain was then applied to analyze the importance of these genera, and the relative importance of variables was visualized using a scree plot. In this plot, variables were arranged in descending order based on their weights, as depicted in Figure 1.

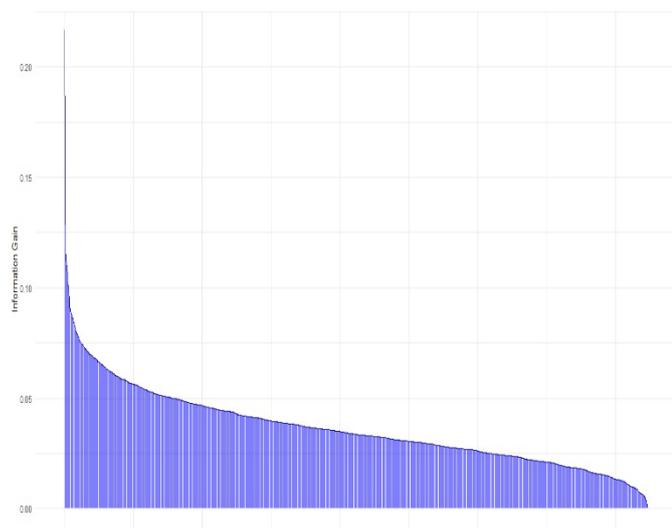

Fig. 1. Information Gain Scree Plot

Upon analyzing the trend in the scree plot depicted in Figure 1, a subset representing 7-8 percent of the variables was selected. Following this, cross-validation was performed to validate the selected features against overfitting. Through this process, the focus narrowed down to 300 bacterial genera considered potentially significant for further analysis.

### 2) LASSO Regression

After identifying these 300 informative bacterial genera related to the target variable, LASSO regression technique was applied to refine the selection and eliminate redundant features. Figure 2 displays the LASSO fit plot, depicting the binomial deviance against the regularization strength (controlled by tuning parameter λ) on the y and x-axes, respectively. This plot assists in pinpointing the optimal λ that balances model complexity and predictive accuracy. The minimum or stabilized point on the plot guides the selection of the most suitable level of regularization for the logistic regression model.

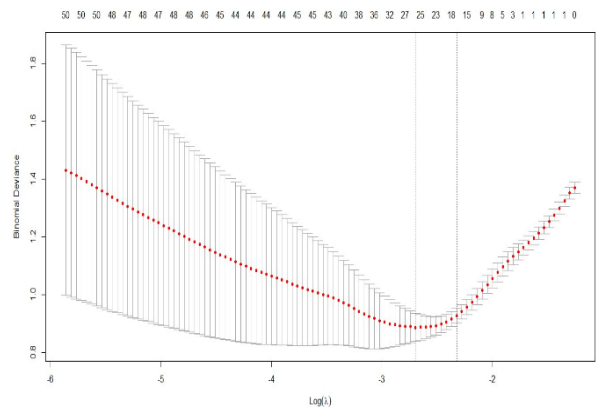

Fig. 2. LASSO Fit plot

Utilizing a lambda LSE value of 0.0979, obtained from Figure 2, the analysis successfully narrowed down the pool to 17 bacterial genera with non-zero coefficients, listed below.

> MPMV01, Aestuariimicrobium, UBA2357, UMGS1071, Harryflintia, NSJ-62, Desulfohalovibrio, Filimonas, Aestuariibacter, UMGS1251, Eggerthella, Blautia, HCH-1, Selenomonas, JAAYOP01, RUG115, Butyribacter

## C. Classification using LDA with LOOCV

After identifying the 17 bacterial genera of interest through LASSO, the ensuing step involved validating the importance of these genera by assessing their prevalence in gastric cancer patients. The validation was carried out using a cross-validated (LOOCV) LDA classifier model. The model demonstrated strong performance across various metrics. It achieved a notable accuracy of 91.67%, with an F1-score of 90.24%. Furthermore, it showed high sensitivity (88%), specificity (94%), and precision (93%). Furthermore, the model demonstrated its robust performance by exhibiting only 8 misclassifications out of 96 participants, highlighting its effectiveness in discriminating between patients with and without gastric cancer.

## V. DISCUSSION

In pursuit of the main objective, which is to distill essential information from a vast reservoir of 10,528 bacterial genera to a focused set of 17, this study navigated the intricate landscape of microbial data to identify key genera holding significant potential in finding patterns for gastric cancer. This reduction, from a multitude to a select few, forms the cornerstone of the investigation, aiming not at prediction but at providing foundational insights.

The investigation commenced with the elimination of non-zero variance, resulting in a significant reduction in the number of bacterial genera. Initially at 10,528 bacterial genera, this number was nearly halved to 5,394, as these genera exhibited zero variance. The pivotal phase of the project involved rigorous feature selection, further paring down the number of bacterial genera to a mere 17 variables. This intricate process included the removal of highly correlated variables, binning, and the implementation of information gain analysis, ultimately identifying the top 300 information-rich genera. Subsequently, the selected 300 genera underwent scrutiny through a LASSO model, resulting in a further reduction to only 17 of the most critical and information-rich bacterium genes. These refined variables now serve as the focal points for the analysis of gastric cancer patients.

Finally, these 17 genera underwent validation through an LDA classifier with the LOOCV. This model demonstrated a remarkable accuracy of 91.67% in effectively classifying patients into healthy or gastrectomy categories based on the identified genera, along with an F1-score of 90.24%. Notably, the model exhibited high specificity (94%), indicating its proficiency in correctly determining true negatives—individuals who do not have gastric cancer identified accurately by the model. Similarly, the model showed high sensitivity (88%), highlighting its capability in correctly identifying true positives—individuals who have gastric cancer identified by the model. Importantly, the model achieved a precision of 93%, confirming its reliability in discriminating between health statuses with only 8 misclassifications out of 96 participants, further affirming its robustness.

The heatmap in Figure 3, utilizing Euclidean distance and complete linkage, shows the normalization values of the bacteria genera of healthy and gastrectomy patients.

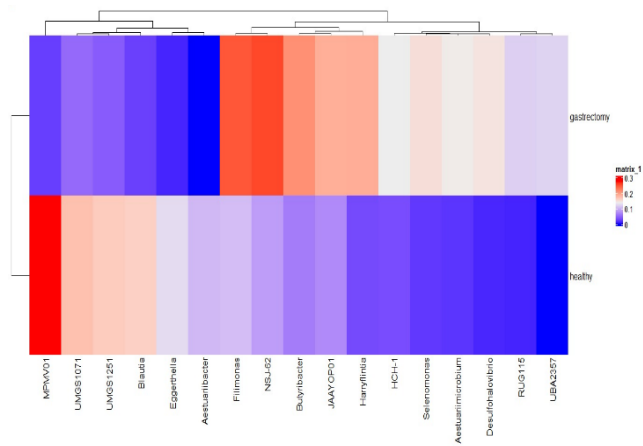

Fig. 3. Heatmap of 17 Genera

A clear distinction between the patients who underwent gastrectomy and those in a healthy state is evident in the heatmap shown in Figure 3.

Although PCA techniques were tried and tested, it must be acknowledged that they were primarily used for visualization purposes. Figure 4 presents a PCA Biplot of the 17 genera, where the red data points correspond to participants who underwent gastrectomy, and the blue data points correspond to healthy participants.

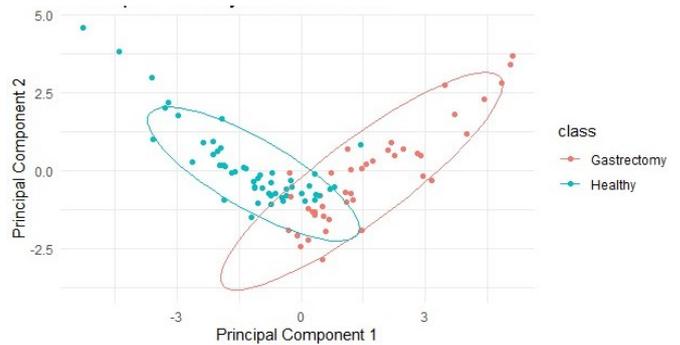

Fig. 4. PCA Biplot

The PCA Biplot in Figure 4 distinctly illustrates the separation between clusters, emphasizing that the 17 genera in the dataset play a significant role in elucidating the outcomes for most patients. This observation underscores the robust explanatory power of the selected genera in distinguishing between different health conditions.

## VI. CONCLUSION

In conclusion, the study successfully narrowed down a vast array of 10,528 bacterial genera to a crucial set of 17, shedding light on the microbial components that hold potential significance in disease discovery. By emphasizing the importance of these selected genera, significant contributions are made not only to the current understanding of the studied pathology but also to paving the way for future investigations in the field. The focus on these 17 genera represents a strategic step towards future endeavors in disease discovery, offering a streamlined approach to comprehending the crucial microbial components associated with the studied pathology.

Moving forward, future work will delve into determining the causal factors of the disease based on the data of the daily habits of patients. This exploration aims to investigate the significance of these habits in the causation of gastric cancer. Furthermore, a comprehensive understanding of the mechanisms through which these subsets of genera impact gastric cancer patients is crucial. This understanding could serve as a foundation for targeted therapy and prevention strategies in the future to assess the risk of a patient who has the likelihood of being affected by gastric cancer depending on the relative concentrations of these bacteria.